\documentclass{article}

\usepackage{arxiv}

\usepackage[utf8]{inputenc} 
\usepackage[T1]{fontenc}    
\usepackage{hyperref}       
\usepackage{url}            
\usepackage{booktabs}       
\usepackage{amsfonts}       
\usepackage{nicefrac}       
\usepackage{microtype}      
\usepackage{graphicx}
\usepackage[numbers]{natbib}
\usepackage{doi}

\usepackage{amssymb}
\usepackage{amsmath} 
\newcommand\tab[1][1cm]{\hspace*{#1}}

\usepackage{tabularx}
\usepackage{booktabs}
\usepackage{multirow}
\usepackage{longtable}

\usepackage{caption}
\usepackage{float}
\usepackage{subcaption}

\title{Emerging Statistical Machine Learning Techniques for Extreme Temperature Forecasting in U.S. Cities}

\author{Kameron B. Kinast \\
	School of Mathematics and Statistics\\
	Rochester Institute of Technology\\
	Rochester, New York 14623 \\
	\texttt{kbk7499@rit.edu} \\
	\And
	Ernest Fokou\'e \\
	School of Mathematics and Statistics\\
	Rochester Institute of Technology\\
	Rochester, New York 14623 \\
	\texttt{epfeqa@rit.edu} \\
}

\date{\today}

\renewcommand{\shorttitle}{\textit{arXiv} Template}

\hypersetup{
pdftitle={A template for the arxiv style},
pdfsubject={q-bio.NC, q-bio.QM},
pdfauthor={David S.~Hippocampus, Elias D.~Striatum},
pdfkeywords={First keyword, Second keyword, More},
}

\begin{document}
\maketitle

\begin{abstract}
	In this paper, we present a comprehensive analysis of extreme temperature patterns using emerging statistical machine learning techniques. Our research focuses on exploring and comparing the effectiveness of various statistical models for climate time series forecasting. The models considered include Auto-Regressive Integrated Moving Average, Exponential Smoothing, Multilayer Perceptrons, and Gaussian Processes. We apply these methods to climate time series data from five most populated U.S. cities, utilizing Python and Julia to demonstrate the role of statistical computing in understanding climate change and its impacts. Our findings highlight the differences between the statistical methods and identify Multilayer Perceptrons as the most effective approach. Additionally, we project extreme temperatures using this best-performing method, up to 2030, and examine whether the temperature changes are greater than zero, thereby testing a hypothesis.
\end{abstract}


\section{Introduction}

For many decades, climate scientists have been concerned about the impact of climate change. Climate change increases the risk to both natural and human systems, and the degree of risk depends on various factors, including extreme hot temperatures \citep{IPCC}. The 2015 Paris Agreement was established to limit the increase in temperatures to 1.5$^{\circ}$C above pre-industrial levels (1850-1900) \citep{Rogelj_den_2016}. Achieving this goal would reduce exposure to climate-related risks, such as heatwaves, droughts, and extreme precipitation events \citep{IPCC}.

Understanding, modeling, and forecasting weather patterns pose ongoing challenges. Time series forecasting is a technique used to predict future observations by analyzing past values. Scientists employ various machine learning and traditional approaches to analyze and predict these events. For example, Kumar and Middey \citep{kumar2023} used a hybrid of random forest and autoregressive integrated moving average (ARIMA) model to project extreme climate indicators. The objective of this paper is to provide a comprehensive review of the effectiveness of commonly used classical and machine learning methods for time series forecasting.

Numerous comparative studies have compared classical and machine learning methods for time series forecasting. For instance, Hill and other colleagues compared the neural network model with classical models such as the Box-Jenkins model, single exponential smoothing model, and naive model, and claimed that the neural network model performed the best\citep{hill_et_al}. Nesreen and other authors \citep{ml_comparison} conducted a comparative study of different machine learning models for time series forecasting and found that multilayer perceptron (MLP) and Gaussian processes (GP) regression were the best models for forecasting with M3 competition data or different types of time series data. Since classical methods still hold significance in time series analysis and forecasting, this paper will study four different time series forecasting methods:

\begin{enumerate}
    \item Auto-Regressive Integrated Moving Average
    \item Exponential Smoothing
    \item Multilayer Perceptron
    \item Gaussian Processes
\end{enumerate}

As mentioned, the Paris Agreement aims to limit the temperature increase to 1.5 degrees Celsius to mitigate the vulnerability to severe climate effects. This research paper investigates the rate of temperature change over a 28-year period (2002-2030). Hypothetically, this paper performs time series forecasting to test the hypothesis that the rate of temperature change is greater than 0 ($\beta_1 > 0$). It is important to note that this research paper analyzes a specific 28-year period, unlike the IPCC's statement, which expresses concerns about temperature increase over the pre-industrial period spanning more than 150 years.

This paper explores various aspects of traditional approaches and machine learning methods for time series forecasting using climate time series data. The research will utilize Python \citep{van1995python} and Julia \citep{bezanson2012julia} to demonstrate the role of statistical computing in understanding climate change. First, an overview of the essential characteristics of the proposed statistical methods is provided. Next, these methods are applied to climate time series data, and the predictive performance of the proposed models is summarized. Finally, the optimal statistical model is used to forecast temperatures for the next seven years, up to 2030.

\section{Statistical Models}

\subsection{Auto-Regressive Integrated Moving Average}

The Auto-Regressive Integrated Moving Average (ARIMA) model is a combination of three model components: the autoregressive (AR) model, the moving average (MA) model, and the integrated method. An autoregressive model of order \textit{p}, denoted as AR(\textit{p}), can be expressed as \cite{Shumway2017}:

\begin{equation} \label{eq:AR}
X_t = \phi_1 X_{t-1} + \phi_2 X_{t-2} + \ldots + \phi_p X_{t-p} + \epsilon_t
\end{equation}

Here, $X_t$ represents the value of the time series, which is defined as a function of past values $X_{t-1},X_{t-2} \ldots X_{t-p}$, along with the constant terms $\phi_p$  and the error term $\epsilon_t$. On the other hand, a moving average model of order \textit{q}, denoted as MA(\textit{q}), can be expressed as:

\begin{equation} \label{eq:MA}
X_t = \theta_1 \epsilon_{t-1} + \theta_2 \epsilon_{t-2} + \ldots + \theta_q \epsilon_{t-q} + \epsilon_t
\end{equation}

In this case, the time series is defined as a function of the errors of past predictions $\epsilon_{t-1}, \epsilon_{t-2}, \ldots, \epsilon_{t-q}$, with $\theta_q$ representing the constant terms. The integrated method is used to deal with non-stationary time series through differencing. The ARIMA model is typically denoted as ARIMA(\textit{p,d,q}), where \textit{p} is the order of the autoregressive model, \textit{q} is the order of the moving average model, and \textit{d} is the order of differencing in the integrated method.

The ARIMA model is based on the Box-Jenkins methodology, which involves an iterative three-step process to develop a forecasting model for the time series \cite{Dimri_Ahmad_Sharif_2020}:

\begin{enumerate}
    \item Ensure the time series is stationary. The Augmented Dickey-Fuller (ADF) test can determine whether the time series is stationary or not.
    \item Identify the model using the autocorrelation function (ACF) and the partial autocorrelation function (PACF). The ACF assesses the correlation between time series considering all the lags or intervals between time periods, while the PACF only considers specific lags.
    \item Estimate the model parameters using goodness-of-fit tests. Model selection is usually based on the Akaike information criterion (AIC) and the Bayesian information criterion (BIC), defined as follows:

                \begin{equation} \label{eq:AIC}
                        AIC = 2k - 2 \ln(L) \mbox{ and } BIC = k \ln(n) - 2 \ln(L)
                \end{equation}
                
    The AIC measures the information value of the model using maximum likelihood estimates, denoted as $L$, and the number of parameters in the model, denoted as $k$. The BIC is similar to the AIC but includes a larger penalty term that takes into account the number of observations, denoted as $n$, in the data.
\end{enumerate}

\subsection{Exponential Smoothing}

Exponential Smoothing (ETS) is a traditional time series approach that forecasts future values based on weighted averages of past observations, assuming that current observations have more weight than past observations. By applying the Holt-Winters additive method, the ETS model with the observed time series $X_t$ can be expressed as \cite{Srivastava_Islam_Singh_Petropoulos_Gupta_Dai_2016}:

\begin{eqnarray}
    Level: L_t = \alpha(X_t - S_{t-p}) + (1- \alpha)(L_{t-1}+ T_{t-1}) \tab  0 < \alpha < 1 \\ 
    Trend: T_t = \beta (L_t - L_{t-1})+ (1-\beta) T_{t-1} \tab  0 < \beta < 1 \\
    Seasonal: S_t = \gamma (X_t - L_t) + (1- \gamma)S_{t-p} \tab  0 < \gamma < 1
\end{eqnarray}

Here, $p$ represents the number of periods in a seasonal cycle (e.g., quarterly, monthly, weekly), and $\alpha$, $\beta$, and $\gamma$ are smoothing parameters. The forecasting equation at time \textit{t} is given by:

\begin{equation} 
    F_{t+n} = L_t + nT_t + S_{t-p+n}, \tab n = 1,2,3,....
\end{equation}

In this equation, $n$ represents the number of time periods ahead of the current time period $t$.

\subsection{Multilayer Perceptrons}

The Multilayer Perceptron (MLP) is a widely used type of feedforward artificial neural network in machine learning applications. The training algorithm for an MLP is known as backpropagation, which involves adjusting the network's weights to minimize the difference between predicted outputs and the actual outputs of the training data. The structure typically consists of an input layer, one or more hidden layers of densely connected neurons with non-linear activation functions, and an output layer. The input layer usually consists of lagged observations of weather variables, while the output layer produces forecasts of future weather conditions.

In this study, the MLP model's structure consists of three layers: the input layer, hidden layer, and output layer. The input layer has three neurons representing three past observations in a sequence, and the output neuron represents the forecasted observation. Let ${\bf x} \in \mathbb{R}^p$ be the input vector, ${\bf z}^{(1)} \in \mathbb{R}^{m_1}$ be the vector of hidden, latent variables for the first hidden layer, and $y \in \mathbb{R}$ be the output. Assuming the MLP has three neurons in the input layer, four hidden neurons in the hidden layer, and one neuron in the output layer, we have:

\begin{eqnarray} \label{eq:MLP}
{\bf z}^{(1)} &=& \sigma({\bf W}^{(1)}{\bf x} + {\bf b}^{(1)})\\ \nonumber
y &=& {\bf W}^{(2)}{\bf z}^{(1)} + {\bf b}^{(2)},
\end{eqnarray}

Here, $\sigma(\cdot)$ is a nonlinear activation function applied element-wise to the output of the linear transformations ${\bf W}^{(1)}{\bf x} + {\bf b}^{(1)}$ and ${\bf W}^{(2)}{\bf z}^{(1)} + {\bf b}^{(2)}$, where ${\bf W}^{(1)} \in \mathbb{R}^{m_1 \times p}$ and ${\bf b}^{(1)} \in \mathbb{R}^{m_1}$ are the weight matrix and bias vector for the first hidden layer, and ${\bf W}^{(2)} \in \mathbb{R}^{m_2 \times m_1}$ and ${\bf b}^{(2)} \in \mathbb{R}^{m_2}$ are the weight matrix and bias vector for the output layer. In this case, the ReLU (Rectified Linear Unit) activation function is used, which sets any negative values to 0 (Equation \ref{eq:RELU}). This function is computationally efficient and provides a simple nonlinear transformation.

 \begin{equation} \label{eq:RELU}
    \sigma(u) = max(0,u) 
\end{equation} 

To prevent overfitting in the MLP model, a regularization technique called dropout is applied. Dropout randomly ignores selected neurons during training, reducing the model's sensitivity to specific neuron weights. The common dropout rate used is $p = 0.2$ in each hidden layer. With the dropout rate represented as $d$, the hidden layer from Equation \ref{eq:MLP} can be described as \cite{JMLR:v15:srivastava14a}:

\begin{eqnarray} 
    d^{(1)} \sim \mbox{Bernoulli}(p), &\\ \nonumber
    \hat{z}^{(1)} = d^{(1)} \cdot z^{(1)} &\\ \nonumber
    z^{(2)} = \phi^{(2)} ( \Sigma W^{(2)} \hat{z}^{(1)} + b^{(2)}).
\end{eqnarray}

This study utilizes the Adaptive Moment Estimation (Adam) optimization method as the optimizer for the MLP model. To estimate the model's losses, a loss function is included, allowing the updating of parameters to minimize the losses. Since the prediction involves real values, this MLP model is considered a regression predictive modeling problem. Therefore, the default loss function, Mean Squared Error (MSE), is used.

\subsection{Gaussian Processes}

Gaussian Processes (GPs) are powerful tools for time series analysis, offering a flexible and non-parametric approach to model complex and non-linear relationships in data. Fundamentally, a GP is a collection of random variables in which all finite-dimensional distributions are joint Gaussian distributions for any finite number \cite{Quadrianto2010}. In the context of time series analysis, a GP can be viewed as a probability distribution over the set of all possible time series functions, with the mean and covariance functions defining the properties of the GP.

The goal of GP is to learn the underlying distribution from the training data, and in order to do that, GP uses the method of Bayesian inference. Assuming that the underlying function generating the time series values at time $t$ is denoted by $f_t$, GP regression models the function $f_t$ as a Gaussian process with a mean function $\mu_t$ and covariance function $\Sigma_t$. Suppose the time series ${X_t}$ is given the observed values up to time $t$, i.e., $\mathcal{D}_t = {X_1, X_2, \ldots, X_t}$, the conditional distribution of $X_{t+1}$ given $\mathcal{D}_t$ is then given by:

The goal of GP is to learn the underlying distribution from the training data, achieved through the method of Bayesian inference. Assuming that the underlying function generating the time series values at time $t$ is denoted as $f_t$, GP regression models the function $f_t$ as a Gaussian process with a mean function $\mu_t$ and a covariance function $\Sigma_t$. Given the observed values of the time series ${X_t}$ up to time $t$, denoted as $\mathcal{D}_t = {X_1, X_2, \ldots, X_t}$, the conditional distribution of $X_{t+1}$ given $\mathcal{D}_t$ is given by:

\begin{eqnarray}
p(X_{t+1} \mid \mathcal{D}_t) &=& \mathcal{N}(\mu_{t+1}, \Sigma_{t+1}),
\end{eqnarray}

The posterior mean and covariance are computed as:

\begin{eqnarray}
\mu_{t+1} &=& k_{t+1}^\top(K_t + \sigma^2 I)^{-1} X_t, \\
\Sigma_{t+1} &=& k(X_{t+1},X_{t+1}) - k_{t+1}^\top(K_t + \sigma^2 I)^{-1} k_{t+1},
\end{eqnarray}

Here, $k_{t+1} = [k(X_1,X_{t+1}), k(X_2,X_{t+1}), \ldots, k(X_t,X_{t+1})]^\top$, $K_t$ is the $t \times t$ matrix with entries $K_{i,j} = k(X_i,X_j)$, $\sigma^2$ is the noise variance, and $I$ is the identity matrix. The kernel functions $k(X_i,X_j)$ specify the correlation between the predicted and observed time series values.

GPs often employ a kernel composition approach where multiple kernels are combined through addition or multiplication to shape the resulting distribution. In this study, the GP kernel composition is expressed as:

\begin{equation}
    k(X_i, X_j) = k_{Noise}(X_i, X_j) + k_{Periodic}(X_i, X_j) + k_{RBF}(X_i, X_j)
\end{equation}

where $k_{Noise}(X_i, X_j)$ represents the noise kernel capturing the noise aspects of the time series, $k_{Periodic}(X_i, X_j)$ represents the periodic kernel capturing seasonal patterns, and $k_{RBF}(X_i, X_j)$ represents the RBF (Radial Basis Function) kernel introducing non-linear trends.

The $k_{Periodic}(X_i, X_j)$ and $k_{RBF}(X_i, X_j)$ are expressed as: 

\begin{eqnarray} \label{eq:periodic}
k_{Periodic}(X_i, X_j) &=& \exp\left(-\frac{2\sin^2\left(\pi|X_i - X_j|/p\right)}{\ell^2}\right)
\end{eqnarray}

and 

\begin{eqnarray} \label{eq:RBF}
k_{RBF}(X_i, X_j) &=& \exp\left(-\frac{(X_i - X_j)^2}{2\ell^2}\right).
\end{eqnarray}

\subsection{Model Selection Criterion}

To determine the optimal time series forecasting model, the performance is evaluated using testing datasets. The performance of the four different time series forecasting models is assessed using the following statistical measures: root mean squared error (RMSE) (Equation \ref{eq:RMSE}) and mean absolute error (MAE) (Equation \ref{eq:MAE}).

\begin{equation} \label{eq:RMSE}
RMSE = \sqrt{\frac{\sum_{i=1}^{N} (y_i - \hat{y}_i)^2}{N}}
\end{equation}

\begin{equation} \label{eq:MAE}
MAE = \frac{1}{N} \sum_{i=1}^{N} |y-\hat{y}|
\end{equation}

In these equations, $y_i$ represents the actual observed values, $\hat{y}_i$ represents the predicted values, and $n$ represents the number of observations in the testing dataset. These statistical measures provide a quantitative assessment of the accuracy and performance of the forecasting models by capturing the differences between the predicted and actual values.

\section{Application to Climate Time Series Data}

\subsection{Data used and study area}

The climate data utilized in this study is obtained from the National Centers for Environmental Information's (NCEI's) Climate Data Online (CDO), a repository managed by the National Oceanographic and Atmospheric Administration (NOAA). CDO provides comprehensive summaries of historical daily land surface observations from locations worldwide.

For this analysis, the climate dataset focuses on two key time series measurements: maximum temperatures and minimum temperatures. Maximum and minimum temperatures are recorded in Fahrenheit. The dataset consists of observations collected from January 1st, 1950, to December 31st, 2022, encompassing data from five airport sites across the United States. The selected airport sites, representing various regions of the United States, are as follows:

\begin{enumerate}
    \item Houston
    \item Chicago 
    \item Boston
    \item San Francisco 
    \item Miami
\end{enumerate}

Given the extensive size of the climate dataset, which spans nearly 75 years of data points for each variable, a mini-batch approach is employed to focus on the most recent twenty years of data. Consequently, the dataset is divided into a training set and a testing set in an 80\% to 20\% ratio, allowing for a comparative analysis of the time series forecasting methods. With this split, the training set comprises approximately 16 years of time series observations, while the testing set encompasses the most recent four years, specifically from 2019 to 2022.

\subsection{ARIMA Results}

The ACF and PACF are critical statistical measures for analyzing time series data. For each dataset, plots of the ACF and PACF are examined to determine the appropriate model order. The selection of the best models is based on the Akaike Information Criterion (AIC) and Bayesian Information Criterion (BIC), and the results are summarized in Table \ref{table:AR_lag}.

\begin{table}[h!]
\centering
\small
\begin{tabular}{ c c c c c c } 
 \hline
 \textbf{Variables} & \textbf{City}  & \textbf{AIC score} & \textbf{BIC score} & \textbf{Model} \\ 
 \hline
  \multirow{5}{4em}{TMAX} & Houston & 3.38 & 20558.26 & ARIMA(88,0,0)\\ 
 & Chicago & 4.04 & 24473.82 & ARIMA(109,0,0)\\
 & Boston & 4.06 & 24590.41 & ARIMA(108,0,0)\\
 & San Francisco & 3.08 &  18412.60 & ARIMA(19,0,0)\\
 & Miami & 2.30 & 14700.06 & ARIMA(100,0,0)\\
 \hline
 \multirow{5}{4em}{TMIN} & Houston & 3.28 & 20070.57 & ARIMA(101,0,0) \\ 
 & Chicago & 3.66 & 22267.02 & ARIMA(106,0,0) \\
 & Boston & 3.23 & 19856.57 & ARIMA(108,0,0)\\
 & San Francisco & 1.95 & 12845.68 & ARIMA(97,0,0)\\
 & Miami & 2.55 & 16098.39 & ARIMA(106,0,0)\\
 \hline
\end{tabular}
\caption{Model Order Selection for ARIMA model of Maximum and Minimum Temperatures among five cities.}
\label{table:AR_lag}
\end{table}

\break

\subsection{ETS Results}

The analysis of the time series data revealed the presence of seasonal components in the maximum and minimum temperatures. Since the training set spans 16 years, it suggests the presence of a recurring pattern within each year. Accordingly, the ETS model, also known as Triple Exponential Smoothing (TES), is employed without considering a trend but with a periodicity of 16 seasons.

\subsection{MLP Results}

Prior to training the MLP model, the training set is split into sequences of three observations, which serve as input neurons for the model. Both the training and testing datasets are divided into batches of 12 samples to update the model's parameters iteratively. The model undergoes 2,000 epochs of training to optimize its parameters and minimize the loss function. \nocite{Lewinson_2020}

\subsection{GP Results}

For the GP model in this study, the kernel composition includes the following components:

\begin{itemize}
    \item a noise kernel with a standard deviation of 1.0
    \item a periodic kernel with a length scale of 0.0, a standard deviation of 1.0, and a periodicity of 1.0
    \item and an RBF kernel with a length scale of 4.0 and standard deviation of 0.0.
\end{itemize}

The parameter values for each kernel were determined through multiple trial-and-error iterations using the root mean squared error (RMSE) as the criterion, as summarized in Table \ref{table:GP_Kernel}. With this fixed kernel composition, the GP model utilizes a zero mean function since it is the standard mean function for time series observation data, implying that the mean value is consistently zero.

\begin{table}[h!]
\small
\centering
\begin{tabular}{ |c|c| } 
 \hline
 Kernel Composition & RMSE Values\\ 
 \hline
 N(1) + P(0,1,1) + RBF(4,0) & 12.51\\
 N(1) + P(0,1,1) & 12.81 \\ 
 P(0,1,1) + RBF(4,0) & 12.81 \\
 N(1) + P(0,1,2) + RBF(4,0) & Failed due to infinite sin values\\ 
 N(1) + P(0,1,0) + RBF(4,0) & Failed due to infinite sin values\\
 N(1) + P(0,1,1) + RBF(2,0) & 12.80 \\ 
 N(1) + P(0,1,1) + RBF(5,0) & 12.83\\
 N(1) + P(0,1,1) + RBF(4,4) & 76\\
 N(1) + P(0,1,0) + RBF(4,4) & 76\\ 
 N(1) + P(0,1,1) + RBF(3,0) & 12.86 \\
 N(1) + P(1,1,1) + RBF(4,0) & Failed due to infinite sin values \\ 
 \hline
\end{tabular}
\caption{Trials and Errors with Values for Kernel in Gaussian Process, using Houston's Maximum Temperatures}
\label{table:GP_Kernel}
\end{table}

\subsection{Comparative Results}

\begin{figure}[H]
     \centering
     \begin{subfigure}[b]{0.4\textwidth}
         \centering
         \includegraphics[width = 6.5cm, height = 6cm]{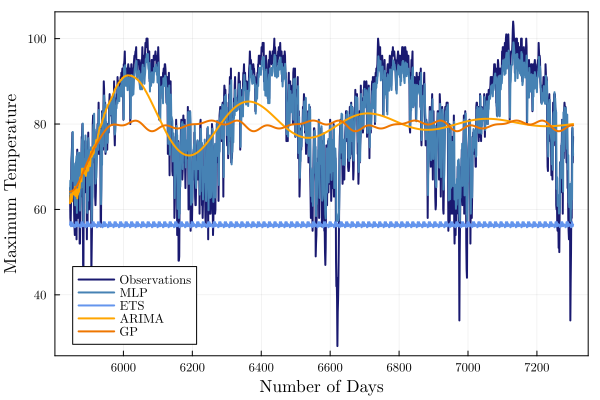}
         \caption{Houston}
     \end{subfigure}
     \hspace{1cm}
     \begin{subfigure}[b]{0.4\textwidth}
         \centering
         \includegraphics[width = 6.5cm, height = 6cm]{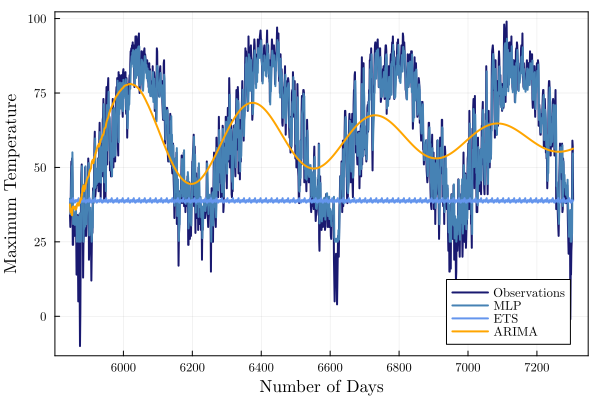}
         \caption{Chicago}
     \end{subfigure}
     \\
         \begin{subfigure}[b]{0.4\textwidth}
         \centering
         \includegraphics[width = 6.5cm, height = 6cm]{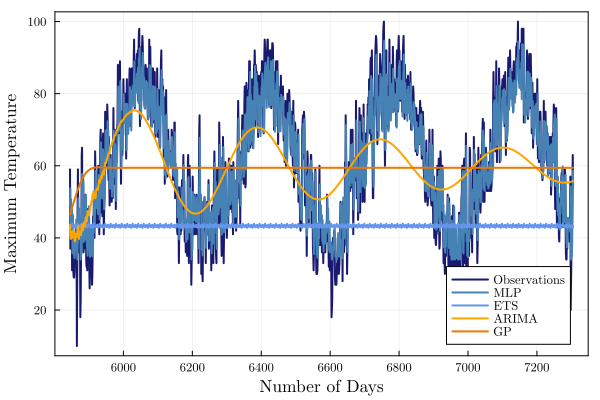}
         \caption{Boston}
     \end{subfigure}
     \hspace{1cm}
         \begin{subfigure}[b]{0.4\textwidth}
         \centering
         \includegraphics[width = 6.5cm, height = 6cm]{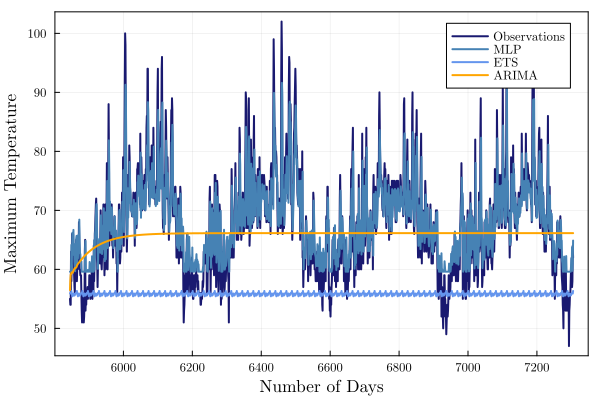}
         \caption{San Francisco}
     \end{subfigure}
     \\
         \begin{subfigure}[b]{0.4\textwidth}
         \centering
         \includegraphics[width = 6.5cm, height = 6cm]{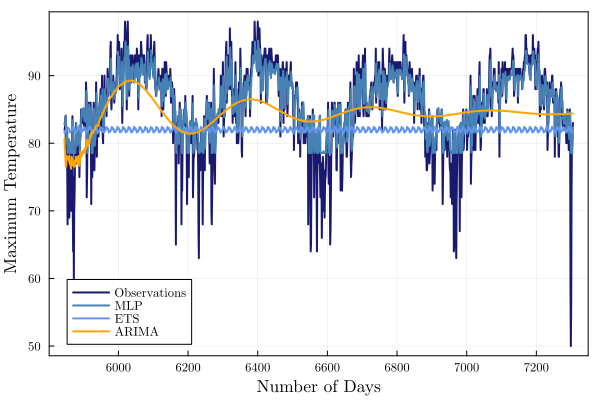}
         \caption{Miami}
     \end{subfigure}
      \caption{Testing set of maximum temperatures values and fitted values predicted by the four methods.}  
    \label{fig:max_plots}
\end{figure}

\begin{figure}[H]
     \centering
     \begin{subfigure}[b]{0.4\textwidth}
         \centering
         \includegraphics[width = 6.5cm, height = 6cm]{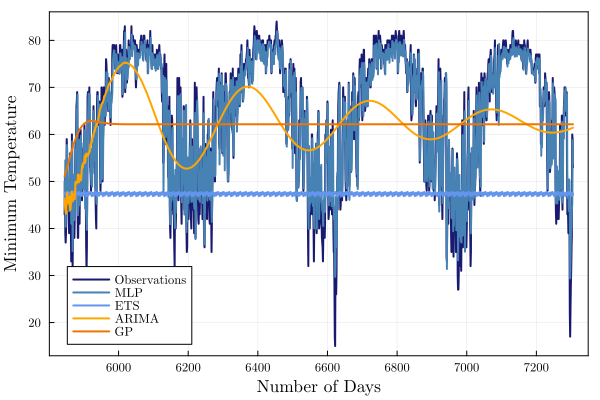}
         \caption{Houston}
     \end{subfigure}
     \hspace{1cm}
     \begin{subfigure}[b]{0.4\textwidth}
         \centering
         \includegraphics[width = 6.5cm, height = 6cm]{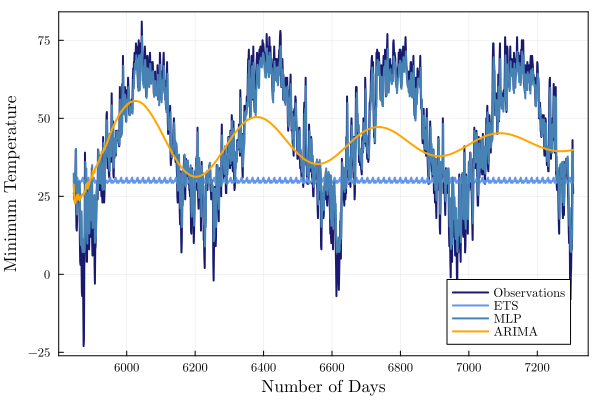}
         \caption{Chicago}
     \end{subfigure}
     \\
         \begin{subfigure}[b]{0.4\textwidth}
         \centering
         \includegraphics[width = 6.5cm, height = 6cm]{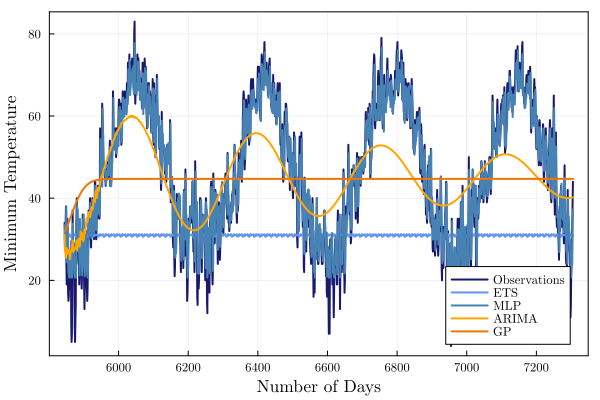}
         \caption{Boston}
     \end{subfigure}
     \hspace{1cm}
         \begin{subfigure}[b]{0.4\textwidth}
         \centering
         \includegraphics[width = 6.5cm, height = 6cm]{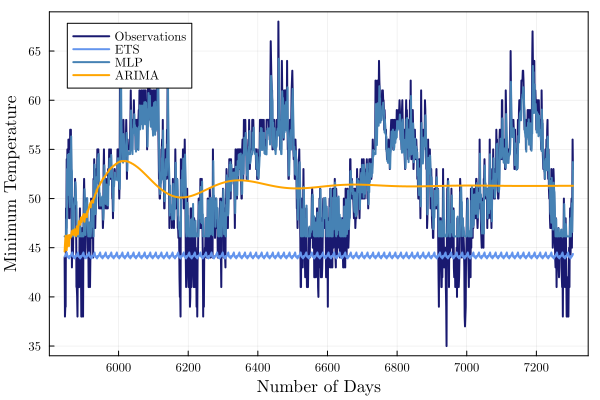}
         \caption{San Francisco}
     \end{subfigure}
     \\
         \begin{subfigure}[b]{0.4\textwidth}
         \centering
         \includegraphics[width = 6.5cm, height = 6cm]{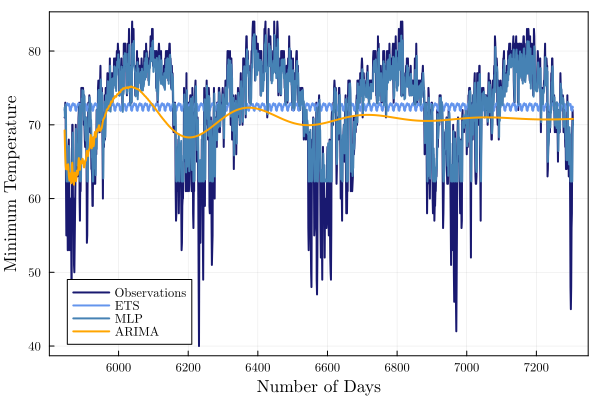}
         \caption{Miami}
     \end{subfigure}
      \caption{Testing set of minimum temperatures values and fitted values predicted by the four methods.}  
    \label{fig:mini_plots}
\end{figure}

The forecasting values of the four methods, along with the testing set of observed values from the NCEI's CDO, are presented in Figure \ref{fig:max_plots} for maximum temperatures and Figure \ref{fig:mini_plots} for minimum temperatures. The graphs illustrate that the predicted values by the MLP model closely match the actual values, unlike the predictions from the other three models.

Table \ref{table:Model_Max} and Table \ref{table:Model_Min} display the model performance of the four methods. Upon examination of the tables, it is observed that the fixed kernel composition of the GP model fails to produce valid predictions for the maximum and minimum temperatures of three cities due to infinite values in the sine function. GP models are known for their instability and ill-conditioning, which means that the fixed kernel composition might be ineffective for these cities, leading to errors in inverse matrices.

Based on the tables, the RMSE and MAE measures indicate that the MLP model outperforms the other three methods. Hence, it can be concluded that the MLP model demonstrates the best performance in forecasting the maximum and minimum temperatures for these five cities. The overall ranking of the four models, based on their performance, is as follows: MLP, ARIMA, ETS, and GP. However, if the model performance is assessed solely based on these two cities, the GP model performs better than the ETS model.

\begin{table}[h]
\centering
\begin{tabular}{c c |c c c c c}
 \hline
 \textbf{Model} & \textbf{Measures} & \textbf{Houston} & \textbf{Chicago} & \textbf{Boston} & \textbf{San Francisco} & \textbf{Miami} \\
 \hline
 \multirow{2}{4em}{ARIMA} & RMSE  & 11.24 & 16.75 & 13.74 & 8.62 & 5.45 \\
 & MAE & 8.90 & 14.10 & 11.33 & 6.69 & 4.29 \\ 
 \hline
 \multirow{2}{4em}{ETS} & RMSE  & 28.14 & 30.28 & 25.54 & 14.62 & 7.11 \\
 & MAE & 25.63 & 25.09 & 20.87 & 12.07 & 5.92\\ 
 \hline
 \multirow{2}{4em}{MLP} & RMSE  & 6.09 & 8.24 & 8.25 & 4.99 & 3.54 \\
 & MAE & 4.42 & 6.33 & 6.53 & 3.81 & 2.29 \\ 
 \hline
 \multirow{2}{4em}{GP} & RMSE  & 12.51 & -- & 17.94 & -- & -- \\
 & MAE & 10.36 & -- & 15.24 & -- & -- \\ 
 \hline
\end{tabular}
\caption{Model Performance Measures for Forecasting on Maximum Temperatures among five cities.}
\label{table:Model_Max}
\end{table}

\begin{table}[h]
\centering
\begin{tabular}{c c |c c c c c}
 \hline
 \textbf{Model} & \textbf{Measures} & \textbf{Houston} & \textbf{Chicago} & \textbf{Boston} & \textbf{San Francisco} & \textbf{Miami} \\
 \hline
 \multirow{2}{4em}{ARIMA} & RMSE  & 11.60 & 16.75 & 11.59 & 5.50 & 6.82 \\
 & MAE & 9.65 & 12.90 & 9.43 & 4.53 & 5.48\\ 
 \hline
 \multirow{2}{4em}{ETS} & RMSE  & 21.50 & 23.67 & 22.52 & 9.41 & 7.32 \\
 & MAE & 18.65 & 19.29 & 18.48 & 8.00 & 5.56\\ 
 \hline
 \multirow{2}{4em}{MLP} & RMSE  & 5.71 & 6.56 & 5.42 & 2.81 & 4.00 \\
 & MAE & 4.02 & 5.01 & 4.07 & 2.13 & 2.80 \\ 
 \hline
 \multirow{2}{4em}{GP} & RMSE  & 13.77 & -- & 16.36 & -- & -- \\
 & MAE & 11.96 & -- & 13.91 & -- & -- \\ 
 \hline
\end{tabular}
\caption{Model Performance Measures for Forecasting on Minimum Temperatures among five cities.}
\label{table:Model_Min}
\end{table}

\vfill

\pagebreak

\pagebreak

\subsection{Temperature Projections}

\begin{figure}[H]
     \centering
     \begin{subfigure}[b]{0.4\textwidth}
         \centering
         \includegraphics[width = 6cm, height = 5cm]{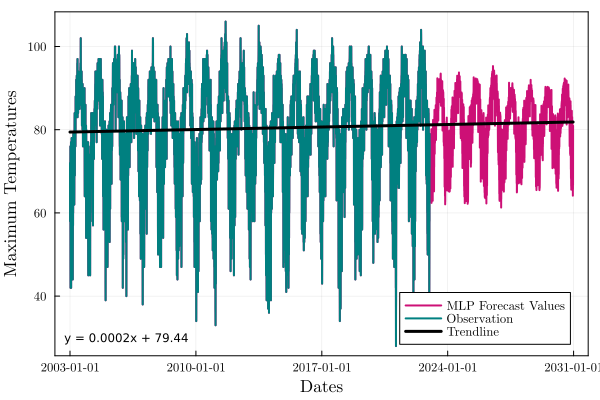}
         \caption{Houston}
     \end{subfigure}
     \hspace{1cm}
     \begin{subfigure}[b]{0.4\textwidth}
         \centering
         \includegraphics[width = 6cm, height = 5cm]{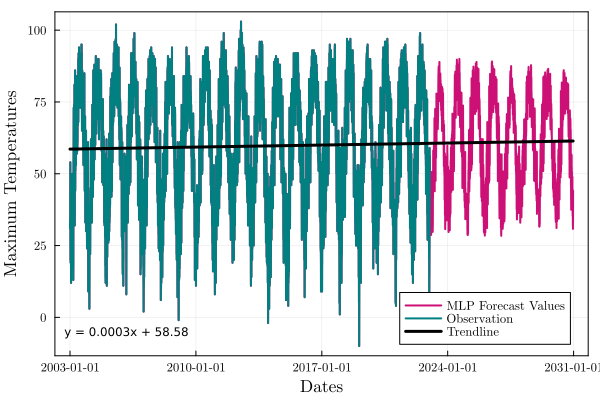}
         \caption{Chicago}
     \end{subfigure}
     \\
         \begin{subfigure}[b]{0.4\textwidth}
         \centering
         \includegraphics[width = 6cm, height = 5cm]{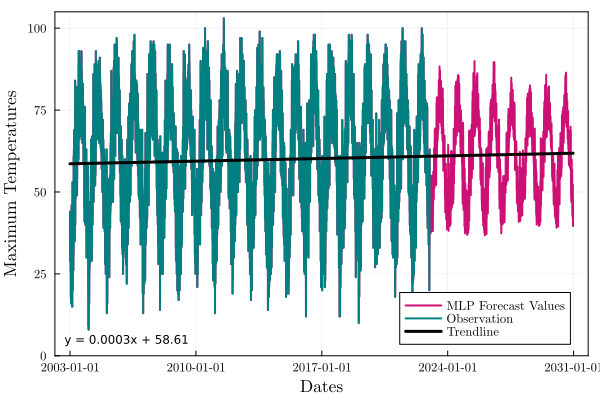}
         \caption{Boston}
     \end{subfigure}
     \hspace{1cm}
         \begin{subfigure}[b]{0.4\textwidth}
         \centering
         \includegraphics[width = 6cm, height = 5cm]{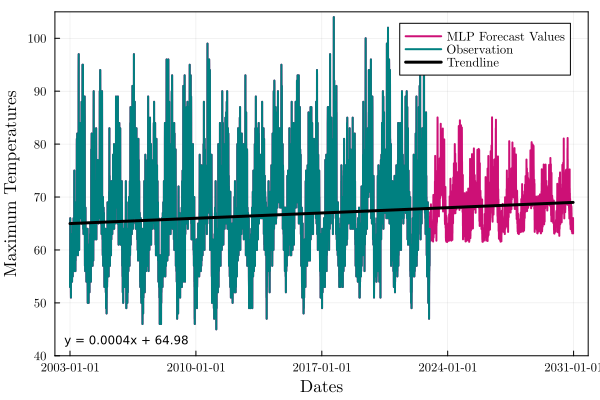}
         \caption{San Francisco}
     \end{subfigure}
     \\
         \begin{subfigure}[b]{0.4\textwidth}
         \centering
         \includegraphics[width = 6cm, height = 5cm]{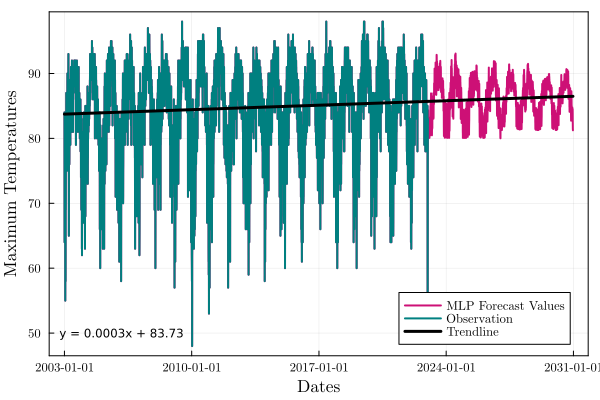}
         \caption{Miami}
     \end{subfigure}
      \caption{Observation values of the maximum temperatures of five cities with the forecast values using MLP method from 2003 to 2030.}  
    \label{fig:max_forecast}
\end{figure}

\begin{figure}[h]
     \centering
     \begin{subfigure}[b]{0.4\textwidth}
         \centering
         \includegraphics[width = 6cm, height = 5cm]{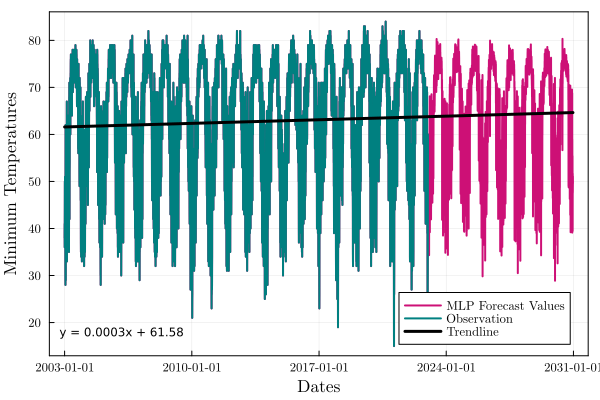}
         \caption{Houston}
     \end{subfigure}
     \hspace{1cm}
     \begin{subfigure}[b]{0.4\textwidth}
         \centering
         \includegraphics[width = 6cm, height = 5cm]{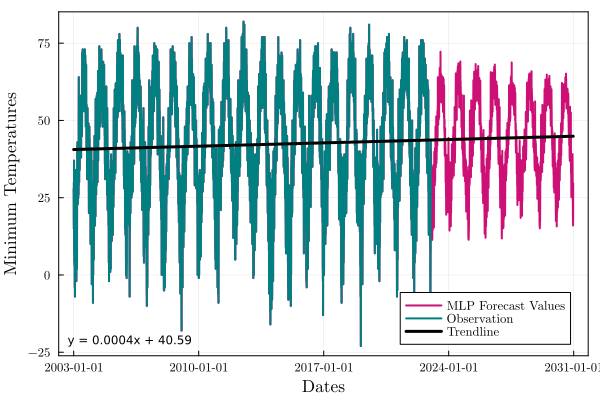}
         \caption{Chicago}
     \end{subfigure}
     \\
         \begin{subfigure}[b]{0.4\textwidth}
         \centering
         \includegraphics[width = 6cm, height = 5cm]{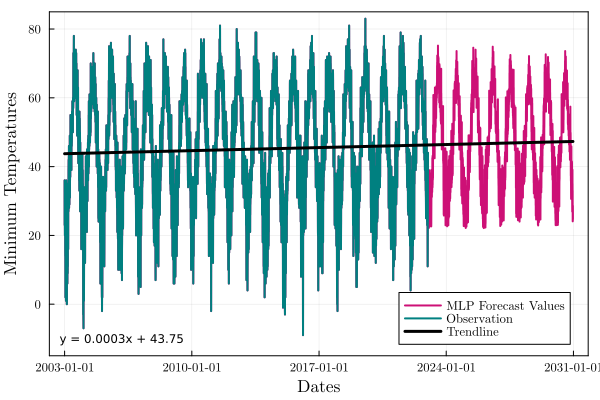}
         \caption{Boston}
     \end{subfigure}
     \hspace{1cm}
         \begin{subfigure}[b]{0.4\textwidth}
         \centering
         \includegraphics[width = 6cm, height = 5cm]{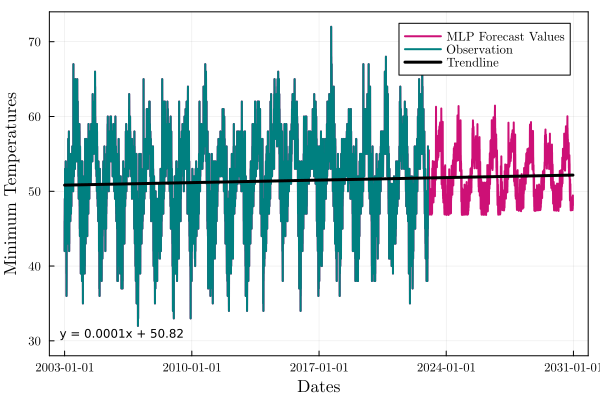}
         \caption{San Francisco}
     \end{subfigure}
     \\
         \begin{subfigure}[b]{0.4\textwidth}
         \centering
         \includegraphics[width = 6cm, height = 5cm]{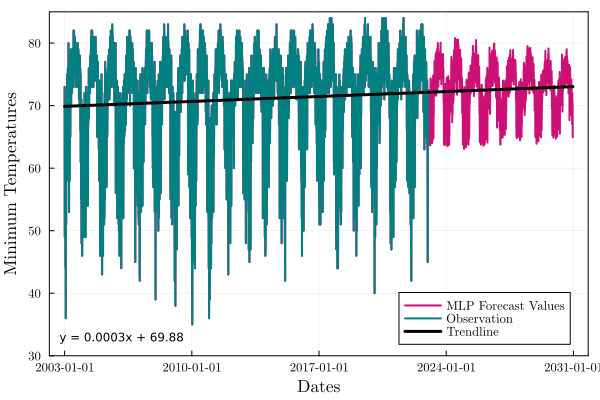}
         \caption{Miami}
     \end{subfigure}
      \caption{Observation values of the minimum temperatures of five cities with the forecast values using MLP method from 2003 to 2030.}  
    \label{fig:mini_forecast}
\end{figure}

Based on the analysis of the figures and the evaluation of model performance, it is evident that the MLP model outperforms the other three statistical models. Figure \ref{fig:max_forecast} presents the projections of maximum temperatures, while Figure \ref{fig:mini_forecast} illustrates the projections of minimum temperatures for all five cities from 2023 to 2030, utilizing the MLP model. The observations values from 2002 to 2022 are incorporated in the model.

Each graph in Figures \ref{fig:max_forecast} and \ref{fig:mini_forecast} includes a regression equation displayed in the bottom left corner. Notably, all regression equations demonstrate that the coefficient $\beta$ is greater than 0, indicating a positive rate of change in the temperatures.

\section{Conclusion}

The foundation of the developed time series methods is based on recent advancements in machine learning and statistical techniques, which provide an optimal framework for each weather forecasting model. In this paper, we explored the feasibility of four different statistical models for time series forecasting of temperatures in five cities. Each city has a unique temperature range, influenced by factors such as its location (coastal or mountainous). We observed that temperatures in all cities exhibit significant seasonal components but no trend.

Based on statistical measures such as RMSE and MAE, the MLP model demonstrates the highest accuracy in predicting maximum and minimum temperatures across all cities. The ARIMA model performs second best, followed by ETS and GP. While this paper suggests that the MLP model is the most effective for time series forecasting of temperatures, it is important to consider certain limitations in model development, such as the focus on only one structure of the MLP model and a fixed kernel composition of GP. Additionally, due to the large dataset size, running these methods on the entire dataset can cause crashes on typical computers. Therefore, this research focuses on a mini-batch of a 20-year dataset instead of 75 years. With access to high-performance computing technology, these machine learning techniques could potentially deliver even better model performance.

The MLP model, as the best model selection, forecasts the temperatures of all five cities up to 2030, and all forecasting performances confirm that the hypothesis is true where there is a positive rate of change, $\beta >0$. Therefore, it indicates that the temperatures, either minimum or maximum, are increasing within only 28 years (2002-2030). However, as presented in the results, the MLP model is not able to forecast extreme temperatures.

It is worth noting that time series analysis encompasses numerous methods beyond the scope of this research, and statistical techniques continue to advance. Furthermore, there are areas for further research expansion. For instance, an empirical comparison study could be extended to include additional machine learning methods for weather time series forecasting, such as Long-Short Term Memory (LSTM), Support Vector Machine (SVM), and Convolutional Neural Network (CNN).

As part of this research, I have created a GitHub repository where all of the code related to my research is stored. The repository is publicly accessible and intended to serve as a resource for anyone interested in the work I have done. By making my code available to the public, I hope to encourage others to build upon my research and to contribute to the wider community of scholars in this field. You can find the repository at \url{https://github.com/kamkinast24/Time-Series-Approach}.

\bibliographystyle{unsrtnat}
\bibliography{references}

\begin{thebibliography}{13}
\providecommand{\natexlab}[1]{#1}
\providecommand{\url}[1]{\texttt{#1}}
\expandafter\ifx\csname urlstyle\endcsname\relax
  \providecommand{\doi}[1]{doi: #1}\else
  \providecommand{\doi}{doi: \begingroup \urlstyle{rm}\Url}\fi

\bibitem[Masson-Delmotte et~al.(2018)Masson-Delmotte, Zhai, Pörtner, Roberts,
  Skea, Shukla, Pirani, Moufouma-Okia, Péan, R.~Pidcock, Matthews, Chen, Zhou,
  Gomis, Lonnoy, Maycock, Tignor, and (eds.)]{IPCC}
V.~Masson-Delmotte, P.~Zhai, H.-O. Pörtner, D.~Roberts, J.~Skea, P.R. Shukla,
  A.~Pirani, W.~Moufouma-Okia, C.~Péan, S.~Connors R.~Pidcock, J.B.R.
  Matthews, Y.~Chen, X.~Zhou, M.I. Gomis, E.~Lonnoy, T.~Maycock, M.~Tignor, and
  T.~Waterfield (eds.).
\newblock Global warming of 1.5°c. an ipcc special report on the impacts of
  global warming of 1.5°c above pre-industrial levels and related global
  greenhouse gas emission pathways, in the context of strengthening the global
  response to the threat of climate change, sustainable development, and
  efforts to eradicate poverty.
\newblock \emph{Summary for Policymakers}, pages 3--24, 2018.
\newblock URL \url{https://doi.org/10.1017/9781009157940.001}.

\bibitem[Rogelj et~al.(2016)Rogelj, den Elzen, Höhne, Fransen, Fekete,
  Winkler, Schaeffer, Sha, Riahi, and Meinshausen]{Rogelj_den_2016}
Joeri Rogelj, Michel den Elzen, Niklas Höhne, Taryn Fransen, Hanna Fekete,
  Harald Winkler, Roberto Schaeffer, Fu~Sha, Keywan Riahi, and Malte
  Meinshausen.
\newblock Paris agreement climate proposals need a boost to keep warming well
  below 2°c.
\newblock \emph{Nature}, 534\penalty0 (7609):\penalty0 631–639, 2016.
\newblock \doi{10.1038/nature18307}.
\newblock URL \url{https://doi.org/10.1038/nature18307}.

\bibitem[Kumar and Middey(2023)]{kumar2023}
N.~Kumar and A.~Middey.
\newblock Extreme climate index estimation and projection in association with
  enviro-meteorological parameters using random forest-arima hybrid model over
  the vidarbha region, india.
\newblock \emph{Environmental Monitoring and Assessment}, 195\penalty0 (380),
  2023.
\newblock \doi{10.1007/s10661-022-10902-2}.
\newblock URL \url{https://doi.org/10.1007/s10661-022-10902-2}.

\bibitem[Hill et~al.(1996)Hill, O'Connor, and Remus]{hill_et_al}
Timothy Hill, Marcus O'Connor, and William Remus.
\newblock Neural network models for time series forecasts.
\newblock \emph{Management Science}, 42:\penalty0 1082--1092, 11 1996.
\newblock \doi{10.1287/mnsc.42.7.1082}.
\newblock URL \url{https://doi.org/10.1287/mnsc.42.7.1082}.

\bibitem[Ahmed et~al.(2010)Ahmed, Atiya, Gayar, and El-Shishiny]{ml_comparison}
Nesreen~K. Ahmed, Amir~F. Atiya, Neamat~El Gayar, and Hisham El-Shishiny.
\newblock An empirical comparison of machine learning models for time series
  forecasting.
\newblock \emph{Econometric Reviews}, 29\penalty0 (5-6):\penalty0 594--621,
  2010.
\newblock \doi{10.1080/07474938.2010.481556}.
\newblock URL \url{https://doi.org/10.1080/07474938.2010.481556}.

\bibitem[Van~Rossum and Drake~Jr(1995)]{van1995python}
Guido Van~Rossum and Fred~L Drake~Jr.
\newblock \emph{Python reference manual}.
\newblock Centrum voor Wiskunde en Informatica Amsterdam, 1995.

\bibitem[Bezanson et~al.(2012)Bezanson, Karpinski, Shah, and
  Edelman]{bezanson2012julia}
Jeff Bezanson, Stefan Karpinski, Viral~B Shah, and Alan Edelman.
\newblock Julia: A fast dynamic language for technical computing.
\newblock \emph{arXiv preprint arXiv:1209.5145}, 2012.

\bibitem[Shumway and Stoffer(2017)]{Shumway2017}
Robert~H. Shumway and David~S. Stoffer.
\newblock \emph{ARIMA Models}, pages 75--163.
\newblock Springer International Publishing, Cham, 2017.
\newblock ISBN 978-3-319-52452-8.
\newblock \doi{10.1007/978-3-319-52452-8_3}.
\newblock URL \url{https://doi.org/10.1007/978-3-319-52452-8_3}.

\bibitem[Dimri et~al.(2020)Dimri, Ahmad, and Sharif]{Dimri_Ahmad_Sharif_2020}
Tripti Dimri, Shamshad Ahmad, and Mohammad Sharif.
\newblock Time series analysis of climate variables using seasonal arima
  approach.
\newblock \emph{Journal of Earth System Science}, 129\penalty0 (1), 06 2020.
\newblock URL \url{https://doi.org/10.1007/s12040-020-01408-x}.

\bibitem[Srivastava et~al.(2016)Srivastava, Islam, Singh, Petropoulos, Gupta,
  and Dai]{Srivastava_Islam_Singh_Petropoulos_Gupta_Dai_2016}
Prashant~K. Srivastava, Tanvir Islam, Sudhir~K. Singh, George~P. Petropoulos,
  Manika Gupta, and Qiang Dai.
\newblock Forecasting arabian sea level rise using exponential smoothing state
  space models and arima from topex and jason satellite radar altimeter data.
\newblock \emph{Meteorological Applications}, 23\penalty0 (4):\penalty0
  633–639, Dec 2016.
\newblock \doi{10.1002/met.1585}.
\newblock URL \url{https://doi.org/10.1002/met.1585}.

\bibitem[Srivastava et~al.(2014)Srivastava, Hinton, Krizhevsky, Sutskever, and
  Salakhutdinov]{JMLR:v15:srivastava14a}
Nitish Srivastava, Geoffrey Hinton, Alex Krizhevsky, Ilya Sutskever, and Ruslan
  Salakhutdinov.
\newblock Dropout: A simple way to prevent neural networks from overfitting.
\newblock \emph{Journal of Machine Learning Research}, 15\penalty0
  (56):\penalty0 1929--1958, 2014.
\newblock URL \url{http://jmlr.org/papers/v15/srivastava14a.html}.

\bibitem[Quadrianto et~al.(2010)Quadrianto, Kersting, and Xu]{Quadrianto2010}
Novi Quadrianto, Kristian Kersting, and Zhao Xu.
\newblock \emph{Gaussian Process}, pages 428--439.
\newblock Springer US, Boston, MA, 2010.
\newblock ISBN 978-0-387-30164-8.
\newblock \doi{10.1007/978-0-387-30164-8_324}.
\newblock URL \url{https://doi.org/10.1007/978-0-387-30164-8_324}.

\bibitem[Lewinson(2020)]{Lewinson_2020}
Eryk Lewinson.
\newblock \emph{Multilayer perceptrons for time series forecasting},
  chapter~10, page 361–376.
\newblock Packt Publishing Ltd., 2020.

\end{thebibliography}

\end{document}